# A Global Inventory of Feedback

Timothy M. Heckman

The William H. Miller III Department of Physics and Astronomy, The Johns Hopkins University, Baltimore, MD 21218, USA

Philip N. Best

Institute for Astronomy, University of Edinburgh, Royal Observatory, Blackford Hill, Edinburgh, EH9 3HJ, UK

**Abstract**

Feedback from both supermassive black holes and massive stars plays a fundamental role in the evolution of galaxies and the inter-galactic medium. In this paper we use available data to estimate the total amount of kinetic energy and momentum created per co-moving volume element over the history of the universe from three sources: massive stars and supernovae, radiation pressure and winds driven by supermassive black holes, and radio jets driven by supermassive black holes. Kinetic energy and momentum injection from jets peaks at $z \approx 1$, while the other two sources peak at $z \approx 2$. Massive stars are the dominant global source of momentum injection. For supermassive black holes, we find that the amount of kinetic energy from jets is about an order-of-magnitude larger than that from winds. We also find that amount of kinetic energy created by massive stars is about 2.5 $\varepsilon_{star}$ times that carried by jets (where $\varepsilon_{star}$ is the fraction of injected energy not lost to radiative cooling). We discuss the implications of these results for the evolution of galaxies and the IGM. Because the ratio of black hole mass to galaxy mass is a steeply increasing function of mass, we show that the relative importance of black hole feedback to stellar feedback likewise increases with mass. We show that there is a trend in the present-day universe which, in the simplest picture, is consistent with galaxies that have been dominated by black hole feedback being generally quenched, while galaxies that have been dominated by stellar feedback are star-forming. We also note that the amount of kinetic energy carried by jets and winds appears sufficient to explain the properties of hot gas in massive halos ($> 10^{13}$ M$_\odot$).

1. Introduction

The basic properties of galaxies, supermassive black holes, and the intra-group/intra-cluster medium cannot be understood without considering the impact of the return of mass, metals, energy, and momentum from both populations of massive stars (stellar winds and supernovae) and supermassive black holes (winds and jets). Examples include the shape of the stellar mass function, the quenching and subsequent suppression of star-formation in massive galaxies, the mass-metallicity and mass-radius relations, the Kennicutt-Schmidt law of star-formation, and the group/cluster X-ray luminosity-temperature relation (see reviews by Somerville & Davé 2015, Naab & Ostriker 2017, Donahue & Voit 2022).

This input of energy and momentum from massive stars and black holes is generically referred to as feedback. Even the highest resolution numerical simulations cannot fully include all the relevant physics *ab-initio,* and must rely on "sub-grid physics" (essentially, recipes for processes that cannot be spatially resolved). The same is true of semi-analytic models. This underscores the importance of using observations to inform the choices that are made in simulations and models. While there is now a considerable body of data on feedback from both massive stars and supermassive back holes (e.g. Veilleux et al. 2020; McNamara & Nulsen 2007; Thompson & Heckman 2023), we still have a very incomplete understanding of the impact of this feedback on the surrounding gas.

In this paper, we will take a different approach from previous investigations of feedback, and try to compile a global inventory (that is, integrated over cosmic time) of the amount of kinetic energy and momentum per co-moving volume element injected by massive stars and supermassive black holes. We will then compare the respective importance of these feedback sources as a function of time and of galaxy and black hole mass.

## 2. Methodology
*2.1 Massive Stars*

To compute the total amount of kinetic energy injected by massive stars (stellar winds and supernovae) per unit volume, we start with the present-day amount of stellar mass per unit volume. We use the compilation in Madau & Dickinson (2016), adjusted to a standard Chabrier Initial Mass Function (IMF - Chabrier, 2003). This value is $3.4 \times 10^8$ $M_\odot$ $Mpc^{-3}$. To compute the corresponding amount of kinetic energy we need to first correct this to account for stars that were formed but are no longer present. This requires multiplication by $1/(1-R)$, where R is the so-called Returned Fraction, which is 0.3 for a Chabrier IMF. Thus, the total mass of stars formed per unit volume is $4.9 \times 10^8$ $M_\odot$ $Mpc^{-3}$. Starburst99 (Leitherer et al. 1999) models for a Chabrier IMF yield a total kinetic energy in stellar winds and supernova ejecta of $6.9 \times 10^{15}$ erg $gm^{-1}$. This then gives a value for the kinetic energy density due to stars of $U_{star} = 6.8 \times 10^{57}$ erg $Mpc^{-3}$.

How much of this kinetic energy is available to supply feedback? The stellar ejecta initially carrying the energy collide and their kinetic energy is converted to thermal energy. This hot gas can then expand and flow outward with the thermal energy being converted back into kinetic energy (e.g. Chevalier & Clegg 1985). Some of the initial thermal energy can be lost through radiative cooling, so that only a fraction $\varepsilon_{star}$ remains to provide feedback. Numerical simulations that represent typical conditions in low-z star-forming galaxies yield $\varepsilon_{star} \approx 0.1$ (Kim et al. 2020), with a value that increases with the star-formation rate per unit area (SFR/A). At the much higher values of SFR/A seen in starbursts (e.g. Kennicutt & Evans 2012), simulations and models predict far greater efficiency, with $\varepsilon_{star} \approx 0.3$ to 1.0 (Schneider et al. 2020; Fielding & Bryan 2022). This is consistent with X-ray observations of the H-like and He-like Fe K$\alpha$ emission-lines in starburst galaxies from the very hot ($10^8$ K) gas created as the stellar ejecta are thermalized through shocks (Thompson & Heckman 2023). These results imply that rather little of the initial kinetic energy is lost through radiative cooling, and this is substantiated by estimates of the rate of PΔV work done by the wind on the ambient gas (Thompson & Heckman 2023). While there are no such constraints on galaxies at high (z > 1) redshift, we do know that these galaxies have values of SFR/A similar to those seen in low-z starbursts (e.g. Forster-Schreiber & Wuyts 2020), and that galactic winds driven by massive stars at this epoch are both ubiquitous and very similar to those seen in low-z starburst galaxies (see Thompson & Heckman 2023). Note that roughly 60% of the total present-day stellar mass was formed at z > 1, during this "windy" epoch (Madau & Dickinson 2016).

The situation for momentum injection is less uncertain because momentum will be conserved even in the face of significant radiative losses. We can simply use the methodology above but use Starburst99 to compute the specific injection rate of momentum by massive stars (supernovae, stellar winds, and radiation pressure). The value is $7.4 \times 10^7$ cm s$^{-1}$, and for a total stellar mass density of $4.8 \times 10^8$ M$_\odot$ Mpc$^{-3}$, this yields $7.1 \times 10^{49}$ gm cm s$^{-1}$ Mpc$^{-3}$.

*2.2 Black-Hole Driven Winds and Radiation Pressure*

Winds driven by supermassive black holes are multi-phase and have been measured in a number of different ways. Molecular outflows have been detected in both emission and absorption (see the review by Veilleux et al. 2020). Calculating kinetic energy outflow rates is conceptually straightforward. The luminosity of a CO transition can be converted into a total molecular gas mass, albeit with uncertainties (Tacconi et al. 2020). The measured outflow velocity and the radius of the outflow then yields a kinetic energy flux given as ½ $M_{gas}$ $v_{out}^3$ $r_{out}^{-1}$. For absorption, the OH column density and outflow velocity yields an outflow rate (for an assumed outflow size and OH/H$_2$ conversion factor). The first compilation of molecular outflows by Fiore et al. (2017) implied typical kinetic energy fluxes of $dE_{wind}/dt \approx 3\%$ $L_{Bol}$, however more recent compilations of measurements (Lutz et al. 2020, Lamperti et al. 2022 and private communication) have yielded much smaller values (median of 0.1%).

Similarly, the outflow rates of warm ionized gas can be measured using the Hα or [OIII]5007 luminosity and measured electron density to derive the total mass of ionized gas and then measuring the outflow velocity and radius of the outflow to determine $dE_{wind}/dt$. Different recent measurements have come to drastically different results, with median values ranging from as high as 1% of $L_{bol}$ (Kakkad et al. 2022) to 0.3 % (Fiore et al. 2017), to 0.1% (Revalski et al. 2021), to 0.01% (Dall'Agnol de Oliveira 2021), to 0.0003% (Trindade Falcao et al. 2021).

An independent measurement of the outflow rate in the warm ionized gas comes from observations of BAL QSOs. Here, the absorption-lines can provide a column density and outflow velocity. Direct measurements of the electron densities can be made using the ratio of column densities in lines arising from an excited state vs. the ground state. Photoionization models using the observed ionizing luminosity (Q) and the inferred value of the ionization parameter (U) then yield a size for the outflow: $r_{out} = (Q/4\pi\ n_e\ c\ U)^{1/2}$ (Miller et al. 2020). With a velocity, radius, and column density, the kinetic energy flux can be estimated. The results span a huge range, from 0.001% to 10% $L_{bol}$ (median value of 0.3%). Highly ionized outflows are also detected in about 40% of AGN (Tombesi et al. 2011) based on X-ray absorption-lines. However, because the size scales of these outflows are so uncertain, the kinetic energy outflow rates are also uncertain (by about two orders-of-magnitude, typically ranging between 0.01 and 1% of $L_{Bol}$ – Tombesi et al. 2012).

It is clear from the above that assigning a value for the ratio of $dE_{wind}/dt$ to $L_{bol}$ is difficult. If we take the median values of 0.1%, 0.3%, and 0.1% $L_{Bol}$ for the molecular, warm-ionized, and highly-ionized phases, we get a total value of 0.5% $L_{Bol}$. Multiplying this by the total bolometric energy density per co-moving volume element volume produced by supermassive black holes of $U_{rad} = 8.6 \times 10^{58}$ erg Mpc$^{-3}$ (Hopkins et al. 2007), yields $U_{wind} = 4.3 \times 10^{56}$ ergs Mpc$^{-3}$. This is 6% as large as the value derived for massive stars. Using the present-day mass per unit volume in supermassive black holes of $\rho_{BH} = 5 \times 10^5$ M$_\odot$ Mpc$^{-3}$ (Hopkins et al. 2007) this wind energy density can also be expressed as $U_{wind} = 5 \times 10^{-4} \rho_{BH} c^2$.

We can also consider the amount of momentum provided by AGN. An initial estimate is implied by the momentum carried by radiation ($U_{rad}/c$) where $U_{rad}$ is the total amount of radiant energy per unit volume produced over cosmic time by AGN. This yields an amount of momentum per unit volume of $2.9 \times 10^{48}$ gm cm s$^{-1}$ Mpc$^{-3}$ (about 4% of the value for massive stars). Since the momentum flux (in the non-relativistic case) is just twice the kinetic energy flux divided by the outflow velocity, we need only consider the momentum carried by the molecular and warm ionized flows (since the BAL QSO and X-ray outflows are over an order-of-magnitude faster, but carry similar kinetic energy fluxes).

For the molecular outflows, the data in Lutz (2020) and Lamperti et al. (2022 and private communication) yield median values of $dp_{wind}/dt$ = 1.0 and 0.7 $L_{Bol}/c$ respectively. The near equality is consistent with the idea that the molecular outflows are driven by radiation pressure. If so, then combining radiation pressure and the molecular outflows would be double-counting in the inventory of momentum.

As noted above, there is a very wide range in the ratio between the kinetic energy flux in the warm ionized gas and the AGN bolometric luminosity, and this translates directly into uncertainties in the ratio of momentum flux and radiation pressure for this gas phase. Estimated median values of this ratio range from ≈10 (Kakkad et al. 2022), to ≈1 (Fiore et al. 2017; Revalski et al. 2021), to ≈0.1 (Dall'Agnol de Oliveira et al. 2021), to ≈0.01 (Trindade Falcao et al. 2021). It appears that the momentum flux in the warm ionized outflows is not likely to be significantly larger than those in the molecular gas or to that carried by radiation.

This represents a total injected momentum per unit volume of at most ≈$10^{49}$ gm cm s$^{-1}$ Mpc$^{-3}$, even if we simply add the three sources (radiation, molecular gas, ionized gas) together. This is still an order of magnitude below the value for massive stars.

*2.3 Black Hole-Driven Jets*

The earliest evidence for the outflow of kinetic energy driven by supermassive black holes came from observations of "double lobes" of synchrotron radio emission that straddled massive elliptical galaxies (Baade & Minkowski 1954). Subsequent radio observations at high angular resolution showed narrow collimated features ("jets") linking the two lobes to the galactic nucleus (see Miley 1980).

It is now possible to quantify the amount of kinetic energy carried by jets as a function of the luminosity of the radio source that they power. This can be done by joint observations of the radio and X-ray emission. The expanding radio sources inflate lobes of relativistic plasma, which in X-rays can be observed as cavities in the surrounding hot gas. Bırzan et al. (2004,2008), Dunn et al. (2005), Rafferty et al. (2006), and Cavagnolo et al. (2010) derived the pΔV work (energy) associated with the cavities in a sample of massive galaxies, groups, and clusters, and used the buoyancy timescale (e.g. Churazov et al. 2001) to estimate their ages. They combined these cavity powers with the monochromatic 1.4 GHz radio luminosities to show that the two were well-correlated. The largest uncertainty in this method is the determination of the cavity energy from the measured pressure and volume: $E_{cav} = f_{cav} p\Delta V$. For the relativistic plasma of the radio lobes the enthalpy of the cavity is $4p\Delta V$. Taking $f_{cav} = 4$, Heckman & Best (2014) derived the following best-fit relation from the cavity data:

1) $dE_{jet}/dt = 1.3 \times 10^{38} (L_{1.4GHz}/10^{26} \text{ W Hz}^{-1})^{0.68}$ W

This empirical relation is very similar to predictions from theoretical models of radio jets. Willett et al. (1999) used synchrotron properties to derive the relation:

2) $dE_{jet}/dt = 2.8 \times 10^{36} (f_W)^{3/2} (L_{1.4GHz}/10^{26} \text{ W Hz}^{-1})^{0.84}$ W

Here $f_W$ is a dimensionless factor (in the range 1 to 20) accounting for the uncertainties in the extrapolation from the population of relativistic electrons that produce the observed radio synchrotron emission to the total energy. Agreement with the X-ray cavity data implies $f_W \approx 10$ to 20 (see Heckman & Best 2014).

We adopt the theoretical relation (equation 2), but calibrated by the cavity data (i.e. taking $f_W$ = 15) and use this to convert the radio luminosity function of AGN between z = 0.1 and 3 (Yuan et al. 2017) into a measure of the evolution in the rate of kinetic energy injection per unit volume by radio jets.

The results are shown in Figure 1, and show that the peak rate of kinetic energy injection by jets occurs at a significantly lower redshift (z ≈ 1) than the peak rate due to massive stars and black-hole-driven winds (z ≈ 2, as also shown in Figure 1). We then integrate the energy injection rate by interpolating the values at z = 0.1, 0.5, 1.0, 2.0, and 3.0 and extrapolating from z = 0.1 to 0 and from z = 3.0 to infinity (this extrapolation does not add significantly to the total - see Figure 1). This then gives a value for the time-integrated total kinetic energy per unit volume due to jets of $U_{jet}$ = 2.6 x $10^{57}$ erg $Mpc^{-3}$. These results are broadly in line with similar estimates derived from low-frequency radio luminosity functions (Kondapally et al., private communication).

The time-integrated kinetic energy input from jets is ≈6 times larger than the value estimated about for black-hole-driven winds, and 40% (400%) the total amount of kinetic energy generated by massive stars for $\varepsilon_{star}$ = 1 (0.1). Alternatively, using the present-day mass per unit volume in supermassive black holes of $\rho_{BH}$ = 5 x $10^5$ $M_\odot$ (Hopkins et al. 2007) this jet energy density can also be expressed as $U_{jet}$ = 2.9 x $10^{-3}$ $\rho_{BH} c^2$.

The kinetic energy carried by jets is in the form of relativistic bulk motion. In this case, the momentum can be taken as p ≈ KE/c. The above value of $U_{jet}$ then implies a momentum density of 8.7 x $10^{46}$ gm cm $s^{-1}$ $Mpc^{-3}$. This is much less than the momentum carried by radiation and winds from supermassive black holes, and the momentum produced by massive stars. Jets are therefore far more important feedback sources in terms of kinetic energy than momentum.

*2.4 The Bottom Line*

For total kinetic energy inventory, the largest single source is either massive stars (for $\varepsilon_{star}$ > 0.4) or jets (for $\varepsilon_{star}$ < 0.4). AGN winds are only important at the <10% level. For the total momentum inventory, massive stars dominate (AGN contribute at the ≈10% level). The peak rate of kinetic energy injection by jets occurs at a substantially lower redshift than that from stars or AGN winds (z ≈ 1 and 2, respectively). These results are summarized in Table 1 and Figure 1.

_________________________________________________________________________

Table 1 – Summary of Feedback Inventory

| 1 | 2 | 3 | 4 | 5 | 6 |
|---|---|---|---|---|---|
| Sample | Log ρ | Log sKE | Log $\rho_{KE}$ | Log sp | $\rho_p$ |
| Massive Stars | 8.69 | -5.11 | 57.83 | 7.87 | 49.85 |
| BH Winds | 5.70 | -3.30 | 56.63 | 10.00 | 49.00 |
| BH Jets | 5.70 | -2.54 | 57.43 | 7.94 | 46.94 |

_________________________________________________________________________

*Notes:*

*Column 2 – The log of the present-day mass density of stars (row 3) and supermassive black holes (rows 4 and 5) formed over cosmic time in units of $M_\odot$ $Mpc^{-3}$.*
*Column 3 – The log of the specific kinetic energy released: energy per unit mass in stars (row 3 and black holes (rows 4 and 5). Given in units of $c^2$, and assuming $\varepsilon_{star}$ = 1.0.*
*Column 4 – The log of the amount of kinetic energy created per unit volume (in ergs $Mpc^{-3}$).*
*Column 5 – The specific momentum created (momentum per unit mass in stars (row 3) and black holes (rows 4 and 5). In units of cm $s^{-1}$.*
*Column 6 – The log of the amount of momentum created per unit volume (gm cm $s^{-1}$ $Mpc^{-3}$).*
_________________________________________________________________________

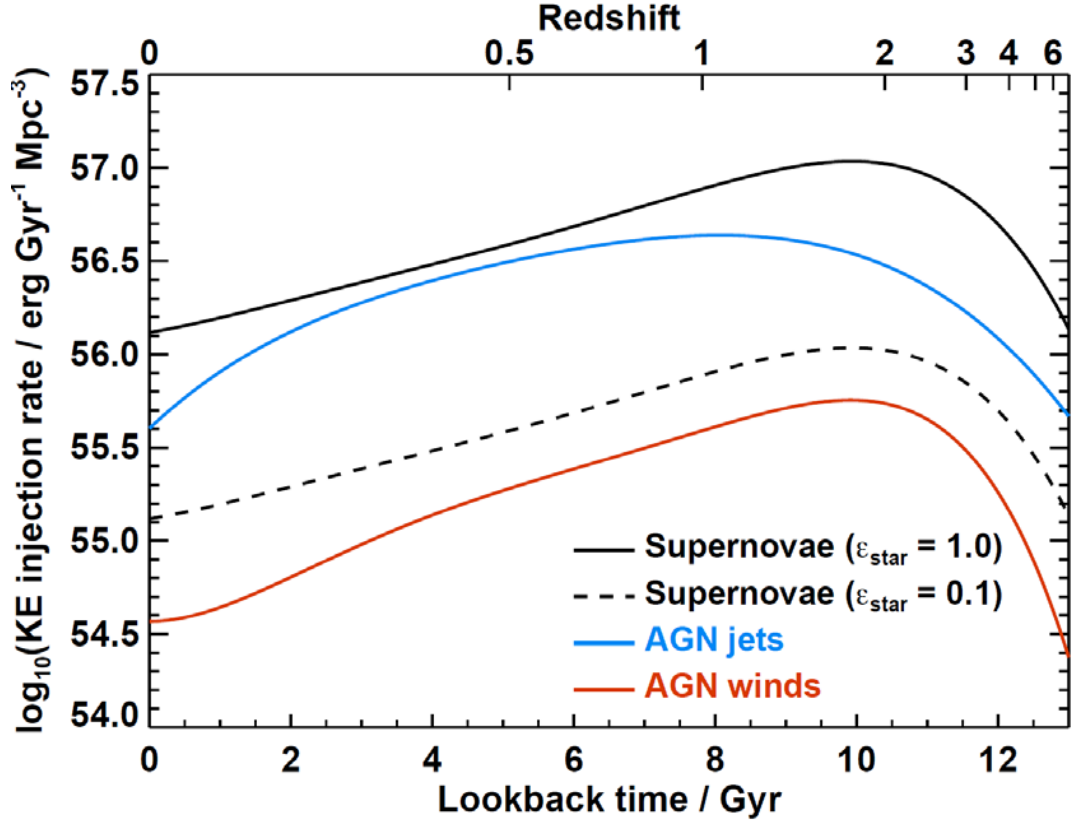

*Figure 1 – A plot of the amount of kinetic energy injected per Gyr and co-moving cubic Mpc as a function of lookback time for massive stars (supernovae and stellar winds; black) and black-hole-driven jets (blue) and winds (red). For massive stars we show the cases in which 100% (solid line) and 10% (dashed line) of the kinetic energy created is delivered to the surroundings (i.e. not lost to radiative cooling). Note that for momentum injection, massive stars dominate at all epochs, with the same time dependence as for kinetic energy injection (i.e. as given by the solid black line).*

### 3. Implications

*3.1 For Galaxies*

To assess the implications of these results for galaxy evolution, it is essential to consider the dependences of feedback on the masses of both galaxies and supermassive black holes. We can go beyond these simple global values and examine the relative importance of feedback (both kinetic energy and momentum) as a function of the ratio of supermassive black hole mass to galaxy stellar mass. In Figure 2 we show a plot of black hole vs. galaxy mass that is similar to that in Heckman & Best (2014) for the $z \approx 0.1$ universe (based on SDSS). In this case, these are present day stellar masses, and would need to be increased by a factor $1/(1-R) = 1.42$ to represent the total mass of stars ever formed. The masses for the black holes were estimated from the M-σ relation from McConnell & Ma (2013). In figure 2, we have color-coded the plot by the fraction of galaxies in which star-formation has been quenched, which we define to be

SFR/M$_{star}$ < 10$^{-11}$ yr$^{-1}$. It is clear that the quenched fraction depends strongly on both the stellar and black hole masses.

The mean relation between stellar and black hole mass in Figure 2 can be approximated as log M$_{BH}$ = 2.0 log M$_{star}$ -14.0, implying M$_{BH}$/M$_{star}$ α M$_{star}$$^{1.0}$ α M$_{BH}$$^{0.50}$. Thus, the relative importance of feedback integrated over cosmic time from massive stars and black holes should be a strong function of mass. Let us quantify this for kinetic energy and then for momentum. For kinetic energy, in a given galaxy (and assuming that global averages can be applied to individual galaxies; see below) the inventories above imply that the ratio KE$_{BH}$/KE$_{star}$ = 315 ε$_{star}$$^{-1}$ M$_{BH}$/M$_{star}$ (where M$_{star}$ is the present-day stellar mass). For momentum, the corresponding ratio is p$_{BH}$/p$_{star}$ = 100 M$_{BH}$/M$_{star}$. We can then plot these relations in Figure 2 to see the regimes in which feedback from supermassive black holes exceeds that from stars. For kinetic energy, we show this separately for values of ε$_{star}$ = 0.1 and 1.0.

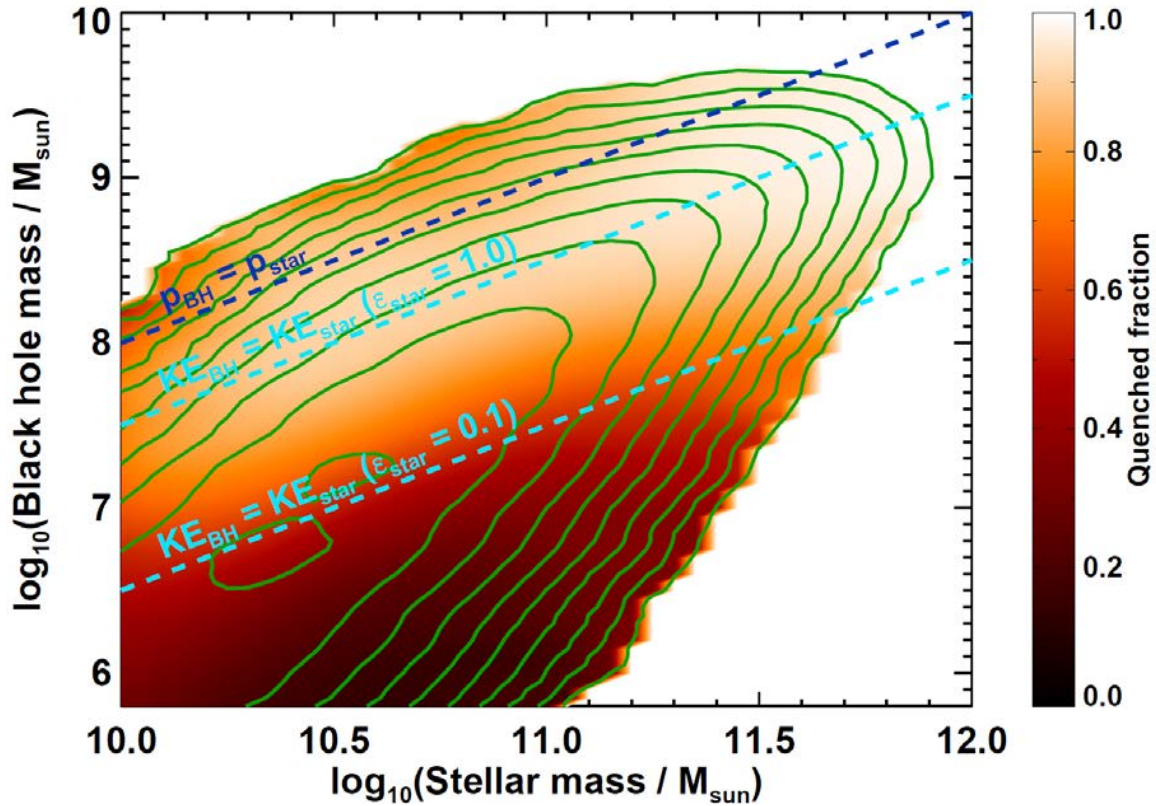

*Figure 2 – A plot of the distribution of SDSS galaxies in the plane of galaxy stellar mass vs. supermassive black hole mass. The latter were estimated using the M$_{BH}$ vs. σ relation in McConnell & Ma (2013). The relative numbers of galaxies in each bin are indicated by the green contours (increasing by factors of 2) and the color-coding represents the fraction of galaxies that are quenched (SFR/M$_{star}$ < 10$^{-11}$ yr$^{-1}$). The dark blue dashed line indicates where the momentum injected by black holes equals that from massive stars. The two light blue dashed lines indicate where the kinetic energy from black holes equals that from massive stars for values of ε$_{star}$ = 0.1 and 1.0 (see text). The transition from predominantly star forming to predominantly quenched galaxies occurs near the relationship for ε$_{star}$ = 0.1.*

In terms of momentum input, stars dominate over black holes in almost all cases. However, considering kinetic energy, we find that the transition from galaxies that are mostly quenched to those that are mostly star-forming occurs very near the dividing line between jet-dominated feedback and stellar dominated feedback for a value of $\varepsilon_{star} \approx 0.1$. This is suggestive evidence that quenching is driven by the feedback of kinetic energy from jets driven by supermassive black holes. However, we caution that the transition from star forming to quiescent galaxies also occurs at the transition from disk dominated to bulge dominated galaxies, so the causal connections between galaxy structure, star formation, and black hole feedback are not entirely clear.

We emphasize that the relations plotted in Figure 2 explicitly assume that the global relations can be applied to individual galaxies, namely that the amount of feedback from massive stars in a given galaxy is proportional to stellar mass and that the amount of feedback from jets is proportional to the black hole mass. The former seems like a safe assumption, but the dependence of the production of radio jets on black hole mass may be complex. We know that there are essentially two populations of radio galaxies (e.g. Heckman & Best 2014). In one case ("radiative mode") the jets are launched by star-forming galaxies and are accompanied by strong nuclear radiation (QSO-like). In the other class ("jet-mode") the jets are launched by quenched galaxies, with little accompanying nuclear radiation. The radiative mode becomes more important at higher luminosities and at higher redshifts. For the jet-mode galaxies, Sabater et al. (2019) find that the probability of producing a jet with a given luminosity depends on both the stellar and black hole mass (and more strongly on the former).

The situation for radiative-mode radio galaxies is less clear, although the indications are that any dependence of the ratio of $KE/M_{BH}$ on $M_{BH}$ or $M_{star}$ is weaker (e.g. Janssen et al. 2012, Kondapally et al. 2022). In the context of Figure 2, it may be that the jet-mode is not the dominant population in terms of actively quenching, since the jet-mode galaxies are already quenched (instead, these may just 'maintain' a quenched state). If quenching is due to jets in radiative-mode galaxies, the dividing line between quenched and star-forming galaxies in Figure 2 would imply that time-integrated amount of jet energy contributed by a radiative mode galaxy is proportional to its black hole mass (i.e. the integrated ratio of jet kinetic energy and energy carried by radiation is independent of black hole mass in these galaxies).

Another way to consider this is to ask how the amount of energy supplied by stars and by black holes scales with the binding energy of the galaxy. We take $E_{bind} \approx M_{star} v_c^2$ where $v_c$ is the galaxy circular velocity. The Tully-Fischer relation for disk galaxies (McGaugh et al. 2000) and the Faber-Jackson relation for ellipticals (Bernardi et al. 2003) both imply $v_c \propto M_{star}^{1/4}$. Thus, we have $E_{bind} \propto M_{star}^{3/2}$. Given that $KE_{star} \propto M_{star}$ and $KE_{BH} \propto M_{BH} \propto M_{star}^2$, this implies that $KE_{star}/E_{bind} \propto M_{star}^{-1/2}$, while $KE_{BH}/E_{bind} \propto M_{star}^{1/2} \propto M_{BH}^{1/4}$. This again underscores the fundamental difference in the mass-dependence of feedback from massive stars and supermassive black holes: feedback from stars becomes increasingly impactful on the galaxy as the mass decreases, while feedback from black holes has greater impact as the mass increases.

*3.2 For the Intra-Group and Intra-Cluster Media*

It has long been known that the basic observed properties of the hot gas in groups and clusters of galaxies ($M_{halo} > 10^{13}$ $M_\odot$) are not consistent with simple models of purely gravitational processes operating during the formation of these systems (see Donahue & Voit 2022 and references therein). A particularly simple example of this is the observed relationship between the X-ray temperature (a proxy for halo mass) and X-ray luminosity. As the halo masses decrease, the observed X-ray luminosities fall further below the relationship expected simply from gravitational infall and heating. These lower luminosities arise because the hot gas in these less-massive halos is more spatially-extended than the dark matter, with the resulting drop in gas density leading to lower X-ray luminosities.

This could be due to the feedback of energy injected into the hot gas, which "lifts" the gas outward. As described above, there is direct observational evidence in the local universe of radio jets delivering energy to the hot gas in groups and clusters. As discussed in Donahue & Voit (2022), for this to be responsible for lifting the hot gas, an amount of kinetic energy equal to ≈0.5% $M_{BH}/c^2$ must be delivered. This is close to the value for jets and AGN winds that we estimated above of ≈0.34%. Note that this could be supplemented by the kinetic energy from massive stars (which would be 0.25 to 2.5 the value for jets for $\varepsilon_{star}$ = 0.1 and 1.0 respectively).

4. **Summary**

Based on a global inventory of the amount of kinetic energy and momentum injected by massive stars (stellar winds and supernovae), and by winds and jets driven by supermassive black holes, we draw the following conclusions:

i) The major sources of kinetic energy are massive stars and jets. Winds driven by supermassive black holes provide <10% of the total. The global ratio of the kinetic energy injected by massive stars to that injected by jets is 2.5 $\varepsilon_{star}$ (where $\varepsilon_{star}$ is the fraction of injected energy from stars that is not lost to radiative cooling).

ii) Massive stars are the dominant source of momentum injection (90% of the total). AGN winds provide 10%, and radio jets are negligible.

iii) The peak in the feedback from jets occurs at z ≈ 1, considerably later than the contributions of AGN-winds and massive stars (peaking at z ≈ 2).

iv) Since the ratio of the mass of the supermassive black hole to the galaxy stellar mass increases steeply with mass, there will be a mass-dependence in the relative importance of feedback from the two sources.

v) For the assumptions that the total amount of kinetic energy from massive stars is proportional to the galaxy's stellar mass, and that the total amount of kinetic energy from a supermassive black hole is proportional to its mass, we find that the populations of quenched and star-forming galaxies occur in the regimes where supermassive black hole feedback and massive star feedback dominate, respectively (for a value of $\varepsilon_{star}$ ≈ 0.1).

vi) By comparing the amount of kinetic energy injected as a function of the binding energy of a galaxy, we show that feedback becomes more impactful as galaxy mass decreases for massive stars, but more impactful as galaxy mass increases for black holes.

vii) The global amount of kinetic energy injected by radio jets and AGN winds per unit volume, combined with the supermassive black hole mass function, yields an efficiency for producing kinetic energy in jets of 0.34% $c^2$. This is very close to the amount of energy needed to explain X-ray luminosity-temperature relation in groups and clusters (0.5% $c^2$).